\documentclass[aps,prl,reprint,superscriptaddress,nofootinbib,floatfix]{revtex4-2}

\usepackage{amsmath,amssymb,bm}
\usepackage{graphicx}
\usepackage{capt-of}
\usepackage{xcolor}
\usepackage[normalem]{ulem}
\usepackage{CJKutf8}
\usepackage{orcidlink}
\usepackage{hyperref}
\hypersetup{hidelinks}

\newcommand{\beq}[1]{\begin{equation}\label{eq:#1}}
\newcommand{\eeq}{\end{equation}}
\newcommand{\bseq}[1]{\begin{subequations}\label{eq:#1}}
\newcommand{\eseq}{\end{subequations}}

\makeatletter
\renewcommand{\p@subsection}{}
\makeatother



\AtBeginDocument{\csname CJK*\endcsname{UTF8}{gbsn}}

\begin{document}

	\title{Deep-Learning-Assisted Bayesian Inference for Extreme-Mass-Ratio Inspirals}

\author{Jian-Jun Song}
\affiliation{Department of Physics, Key Laboratory of Low Dimensional Quantum Structures and Quantum Control of Ministry of Education, and Hunan Research Center of the Basic Discipline for Quantum Effects and Quantum Technologies, Hunan Normal University, Changsha 410081, China}

\author{Zhong-Wu Xia}
\affiliation{Department of Physics, Key Laboratory of Low Dimensional Quantum Structures and Quantum Control of Ministry of Education, and Hunan Research Center of the Basic Discipline for Quantum Effects and Quantum Technologies, Hunan Normal University, Changsha 410081, China}

\author{Wen-Hong Ruan}
\email{ruanwenhong@hunnu.edu.cn}
\affiliation{Department of Physics, Key Laboratory of Low Dimensional Quantum Structures and Quantum Control of Ministry of Education, and Hunan Research Center of the Basic Discipline for Quantum Effects and Quantum Technologies, Hunan Normal University, Changsha 410081, China}

\author{{Zu-Cheng Chen}\orcidlink{0000-0001-7016-9934}}
\email{zuchengchen@hunnu.edu.cn}
\affiliation{Department of Physics, Key Laboratory of Low Dimensional Quantum Structures and Quantum Control of Ministry of Education, and Hunan Research Center of the Basic Discipline for Quantum Effects and Quantum Technologies, Hunan Normal University, Changsha 410081, China}

\author{{Hongwei Yu}
\orcidlink{0000-0002-3303-9724}}
\email{hwyu@hunnu.edu.cn}
\affiliation{Department of Physics, Key Laboratory of Low Dimensional Quantum Structures and Quantum Control of Ministry of Education, and Hunan Research Center of the Basic Discipline for Quantum Effects and Quantum Technologies, Hunan Normal University, Changsha 410081, China}

\author{{Puxun Wu}
	\orcidlink{0000-0002-9188-7393}}
\email{pxwu@hunnu.edu.cn}
\affiliation{Department of Physics, Key Laboratory of Low Dimensional Quantum Structures and Quantum Control of Ministry of Education, and Hunan Research Center of the Basic Discipline for Quantum Effects and Quantum Technologies, Hunan Normal University, Changsha 410081, China}

\begin{abstract}

Extreme-mass-ratio inspirals (EMRIs) pose a major challenge to Bayesian parameter inference because of their complex, highly multimodal likelihood surfaces. Locating narrow high-likelihood regions is particularly difficult without knowledge of the true source parameters, which is unavailable in real observations. We develop a deep-learning-assisted Bayesian inference framework that
combines neural-network-guided parameter-space localization, particle swarm optimization, and Markov chain Monte Carlo~(MCMC) sampling. Applied to simulated six-month EMRI signals, the framework recovers posterior distributions in close agreement with reference MCMC analyses initialized near the true parameter values, without requiring such knowledge for initialization. Once the neural network has been trained, the complete inference pipeline takes approximately 10 hours per signal.

\end{abstract}

	\maketitle

	\begin{table*}[!t]
    \centering
    \caption{
        Prior distributions of the source parameters used
        for neural network training and MCMC sampling.
        Here $\mathcal{U}$ denotes a uniform distribution.
    }
    \label{tab:parameter_summary}
    \small
    \setlength{\tabcolsep}{5pt}
    \renewcommand{\arraystretch}{1.15}
    \begin{tabular*}{\textwidth}{@{\extracolsep{\fill}}lp{0.43\textwidth}c}
        \hline\hline
        Parameter & Description & Prior \\
        \hline
        $\ln(M/M_\odot)$ & MBH mass
            & $\mathcal{U}[\ln(8.50\times10^5),\,\ln(1.15\times10^6)]$ \\
        $\ln(\mu/M_\odot)$ & Secondary compact object mass
            & $\mathcal{U}[\ln(9.00),\,\ln(11.00)]$ \\
        $a$ & Dimensionless spin of the MBH
            & $\mathcal{U}[0.8,\,0.99]$ \\
        $p_0\,[M]$ & Initial semilatus rectum
            & $\mathcal{U}[8.5,\,9.5]$ \\
        $e_0$ & Initial orbital eccentricity
            & $\mathcal{U}[0.1,\,0.3]$ \\
        $Y_0$ & Cosine of the initial orbital inclination
            & $\mathcal{U}[0.59,\,0.79]$ \\
        $\cos\theta_S$ & Cosine of the source polar angle
            & $\mathcal{U}[-0.99999,\,0.99999]$ \\
        $\phi_S\,[\mathrm{rad}]$ & Source azimuthal angle
            & $\mathcal{U}[0,\,2\pi]$ \\
        $\cos\theta_K$ & Cosine of the spin polar angle of the MBH
            & $\mathcal{U}[-0.99999,\,0.99999]$ \\
        $\phi_K\,[\mathrm{rad}]$ & Spin azimuthal angle of the MBH
            & $\mathcal{U}[0,\,2\pi]$ \\
        $d_L\,[\mathrm{Gpc}]$ & Luminosity distance
            & $\mathcal{U}[0.01,\,10]$ \\
        \hline\hline
    \end{tabular*}
\end{table*}

	\begin{figure*}[!t]
		\centering
		\includegraphics[width=0.96\textwidth]{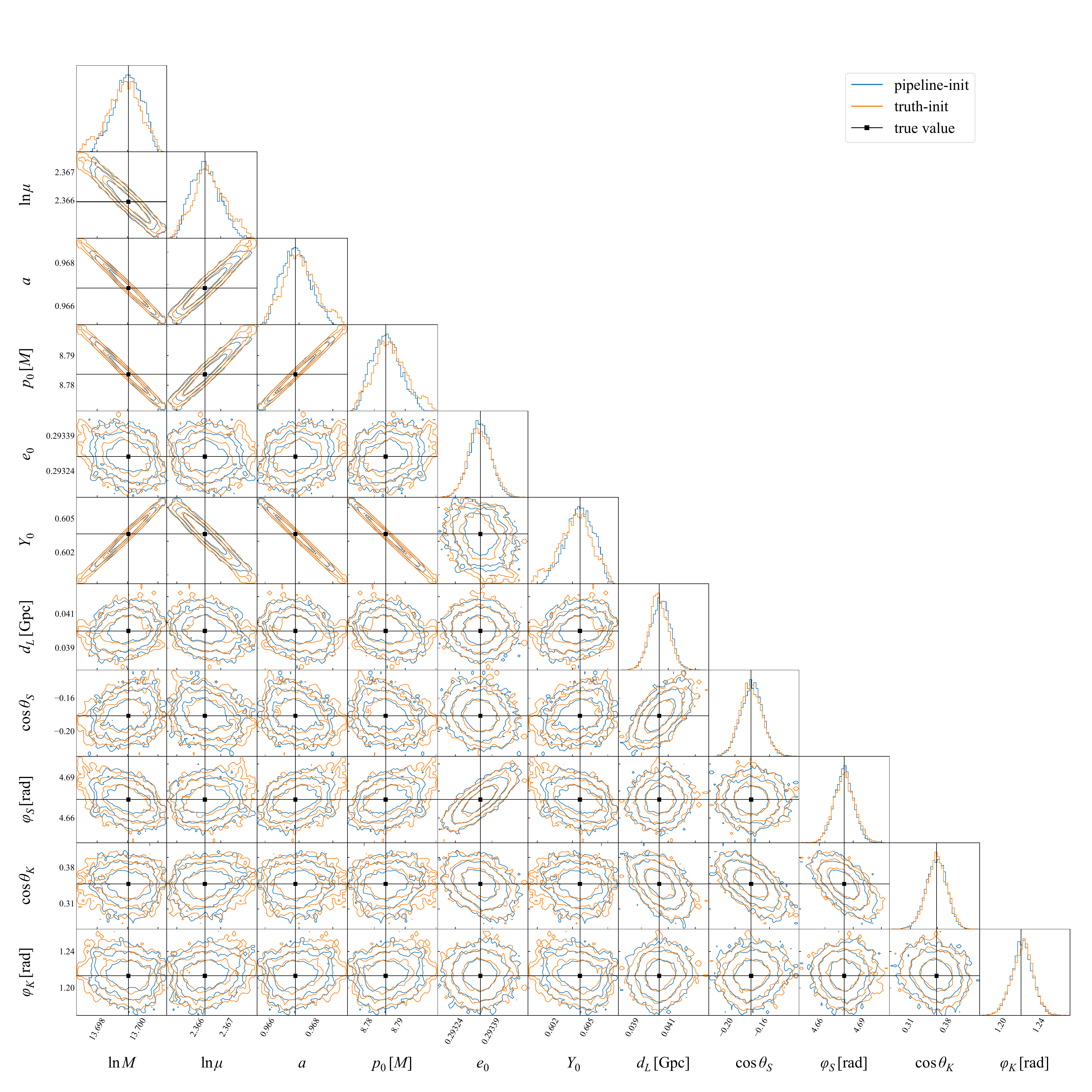}
		\caption{
            Comparison of the posteriors for Signal~1 using
            $6$ months of data, with a signal-to-noise ratio (SNR) of $95.94$.
            Blue shows the posterior obtained through MCMC analyses with
            progressively longer data segments, with the chains in the first stage
            initialized near the parameter values identified by the PSO search assisted by the
            neural network.
            Orange shows the posterior obtained by initializing MCMC chains
            near the true parameter values.
            The contours show the $68.27\%$, $95.45\%$, and $99.73\%$
            credible regions.
            Black lines and markers indicate the true values.
            Both analyses use the same priors and MCMC settings and yield
            posteriors in close agreement.
        }
		\label{fig:result_high_snr}
	\end{figure*}

	\paragraph*{Introduction---}

Extreme-mass-ratio inspirals (EMRIs) are systems in which a
stellar-mass compact object gradually inspirals into a massive
black hole (MBH)~\cite{Amaro-Seoane:2012lgq,Babak:2017tow,Berry:2019wgg,Zenginoglu:2011zz}.
They are important gravitational-wave (GW) sources for future
detectors in space, including the Laser Interferometer Space
Antenna (LISA)~\cite{LISA:2017pwj,LISA:2022yao},
Taiji~\cite{Hu:2017mde,Luo:2021qji,Gong:2021gvw}
and TianQin~\cite{TianQin:2015yph,TianQin:2020hid,Gong:2021gvw}.
EMRIs can complete tens of thousands of orbits while emitting GWs
in the millihertz band, with a substantial fraction of their evolution occurring in
the strong-field region near the MBH~\cite{Amaro-Seoane:2012lgq,Babak:2017tow,Berry:2019wgg}.
Their signals provide valuable probes of  stellar dynamics
in galactic nuclei and MBH populations~\cite{Barack:2003fp,Amaro-Seoane:2007osp,Babak:2017tow,Berry:2019wgg}, as well as the environments surrounding
MBHs~\cite{Barausse:2014tra}, including dark matter distributions~\cite{Cole:2022yzw,Cardoso:2022whc,Dyson:2025dlj,Vicente:2025gsg,Eda:2013gg,Duque:2023seg} and accretion disks~\cite{Speri:2022upm,Yunes:2011ws}. EMRIs also offer opportunities  to test gravity in the strong-field regime~\cite{Ryan:1995wh,Barack:2006pq} and search for new fundamental fields~\cite{Maselli:2021men,Speri:2024qak,Zi:2026zpw,Maselli:2020zgv,Baumann:2022pkl}.

Realizing this scientific potential requires accurate inference of source parameters  from GW observations. Although Bayesian inference based on matched filtering is widely used for GW parameter estimation~\mbox{\cite{Thrane:2018qnx}}\hskip0pt, its application to EMRIs is computationally challenging because repeated likelihood evaluations require generating waveforms that span months or even years~\cite{Liang:2025izu}.
The problem  is further complicated  by the highly multimodal  likelihood surfaces and nonlocal  degeneracies among source parameters,  making it difficult to locate high-likelihood regions without accurate initial parameter estimates~\cite{Gair:2008zc,Wang:2012xh,Chua:2018yng, Chua:2021aah,LISA:2022yao,Cole:2025sqo,Gair:2004iv,Ye:2023lok}. To facilitate posterior sampling, analyses of simulated EMRI signals often adopt narrow priors centered on the true parameter values or initialize Markov chain Monte Carlo (MCMC) chains near them~\cite{Xia:2026aty,Katz:2021yft,Speri:2022upm,Speri:2024qak,Cole:2025sqo,Khalvati:2024tzz,Long:2026dcb}. However, such strategies rely on knowledge of the true source parameters,  which is  unavailable in real GW observations.

Deep learning offers a promising alternative for GW parameter inference. By learning the posterior distributions of source parameters conditioned on GW signals, neural networks can generate posterior samples rapidly. Such methods have been explored extensively for compact binary coalescence signals~\cite{Gabbard:2019rde,Green:2020hst,Green:2020dnx,
Dax:2021tsq,Dax:2022pxd,Langendorff:2022fzq,Dax:2024mcn,
Kofler:2025dux}, MBH binaries~\cite{Chua:2018woh,Chua:2019wwt,Dax:2021tsq,Dax:2022pxd,Ruan:2024qch,Spadaro:2026mbh},
stochastic GW backgrounds~\cite{Shih:2023jme,Vallisneri:2024xfk,Alvey:2023npw,
Alvey:2024uoc,Einsle:2025xsh}, and EMRI signals~\cite{Liang:2025ovq,Liang:2025izu,Cole:2025sqo,Shen:2026zkv,Speri:2025ucn}.  

Despite these advances, accurately approximating posterior distributions for EMRI signals using neural networks over broad prior ranges remains challenging.  The long duration and rich harmonic structure of EMRI signals~\cite{Drasco:2005kz} lead to high-dimensional inputs, making it difficult to reduce the data dimensionality without losing waveform information essential for  training the network~\cite{Cole:2025sqo}.
In addition, the joint posterior distribution of source parameters can exhibit pronounced multimodality and strong parameter correlations, making it challenging for neural networks to accurately capture its complex structure~\cite{Green:2020hst} and potentially increasing the computational cost of training~\cite{Dax:2021tsq,Roulet:2026labrador}.

An alternative strategy is to  employ global optimization to identify promising  high-likelihood regions in the parameter space~\cite{Cornish:2008zd,Zou:2024jqv,Zou:2024osb} before performing  MCMC sampling~\cite{Strub:2025dfs}. However, a direct search over a large parameter space can itself  be computationally expensive, particularly for EMRIs on generic orbits around spinning MBHs, where coherent likelihood evaluations require costly waveform generation~\cite{Zou:2026hal}. This motivates an approach that uses neural networks to guide the search without relying on their approximate posteriors for the final parameter estimates.

In this Letter, we develop a  deep-learning-assisted Bayesian inference framework for EMRIs  that integrates neural-network-guided parameter-space localization, particle swarm optimization (PSO), and MCMC sampling.
The neural network identifies reduced search regions and provides candidate parameter values for initializing
PSO, which subsequently locates suitable starting points for staged MCMC sampling.
 Tests on $20{,}000$ simulated EMRI signals show substantial reductions in the parameter search volume
while maintaining high coverage of the true values.   Applied to six-month EMRI signals, the framework yields posterior distributions in close agreement with reference MCMC analyses initialized near the true parameter values. Once the network has been trained, the complete inference procedure takes approximately $10\,\mathrm{h}$ per signal. By combining  neural-network-assisted localization and likelihood-based posterior sampling, our framework provides a practical approach to EMRI parameter inference without requiring knowledge of the true source parameters for initialization.

\paragraph*{Method: EMRI Waveforms and Data Preprocessing---}
For the training data, EMRI waveform polarizations are generated
using the augmented analytic kludge (AAK) model~\cite{Chua:2017ujo} implemented in
\texttt{FastEMRIWaveforms} (FEW)~\cite{Chua:2020stf,Katz:2021yft,Speri:2024qak}.
The LISA detector response is then computed with
\texttt{FastLISAResponse} to obtain the $A$ and $E$ channels of
second-generation time-delay interferometry (TDI)~\cite{Katz:2022yqe,Tinto:2022zmf}.
This waveform is specified by $14$ source parameters.
Of these, $11$ are drawn independently from the priors listed in
Table~\ref{tab:parameter_summary}, while the three initial phases are fixed to zero.
The training data consist of these source parameter sets and their
associated data $\bm d=(\bm d_A,\bm d_E)$.
The corresponding data are generated for each parameter set with an
observation time of $T_{\rm obs}=3$ months and a sampling interval
of $\Delta t=10$ s.
As the present work primarily focuses on the development
and assessment of an inference methodology, all simulated
data are generated without added noise.

Before being fed into the neural network, the data are preprocessed
to facilitate training.
Specifically, we normalize the data in the $A$ and $E$ channels
separately by their respective root-mean-square (RMS) amplitudes,
\beq{rms}
R_I=
\left[
\frac{1}{N}\sum_{n=1}^{N} \bigl(\bm{d}_I[n]\bigr)^2
\right]^{1/2},
\qquad I=A,E,
\eeq
where $N$ is the number of samples in each channel and
$\bm{d}_I[n]$ denotes the $n$th sample in channel $I$.
Then, the normalized data are transformed into multiresolution
time-frequency representations using short-time Fourier transforms
(STFTs)~\cite{Marulanda:2020nww,Du:2025fes}. We set the fast Fourier transform (FFT) lengths to
$16{,}384$, $32{,}768$, and $65{,}536$ for the three STFT resolutions.
For each resolution, the window length is set equal to the
corresponding FFT length, with a Hann window and $75\%$
overlap between adjacent windows.
To reduce the input dimensionality, average pooling is applied in two
dimensions to obtain a $16\times64$ representation at each resolution.
These multiresolution representations, together with
$\log R_A$ and $\log R_E$, form the input to the embedding
network.

\paragraph*{Bayesian Inference Framework---}
\label{sec:ml}
Our Bayesian inference framework has three main components: a neural network, PSO~\cite{Kennedy:1995mri}, and MCMC. The neural network consists of an embedding network and a neural spline flow (NSF)~\cite{Rezende:2015ocs,Durkan:2019nsq}. The embedding network encodes the information in the data into a compact feature representation by applying a temporal convolutional network~\cite{Bai:2018tcn} and a multilayer perceptron~\cite{Goodfellow:2016dl}. These features are then passed to the NSF, which models the conditional distribution $q(\bm{\theta}|\bm{d})$ of the EMRI source parameters. The NSF consists of 22 rational quadratic spline coupling transforms, each parameterized by a residual network~\cite{He:2016resnet} with 10 residual blocks. The network is trained by minimizing  the cross-entropy loss~\cite{Goodfellow:2016dl,Green:2020dnx}.

Once trained,  the neural network rapidly generates approximate posterior samples for  a given signal. The $99.73\%$ marginal credible intervals inferred from these samples are then used to define a reduced, axis-aligned search region.  A subset of the PSO particles is initialized using network-generated samples within this region (see Supplemental Material~\cite{SupplementalMaterial} for details of the search region construction).

PSO then searches the reduced  region for parameter values
that maximize the joint $A/E$ overlap $\mathcal O_{AE}$,
thereby identifying a suitable starting point for subsequent  MCMC sampling. Here, $\mathcal O_{AE}$ is defined as~\cite{Katz:2021yft,Zou:2024jqv}
\beq{pso_overlap}
\mathcal{O}_{AE}(\bm d,\bm h)
=
\frac{
\bm d_A\!\cdot\!\bm h_A+
\bm d_E\!\cdot\!\bm h_E
}{
\sqrt{
\left(
\bm d_A\!\cdot\!\bm d_A+
\bm d_E\!\cdot\!\bm d_E
\right)
\left(
\bm h_A\!\cdot\!\bm h_A+
\bm h_E\!\cdot\!\bm h_E
\right)
}
}.
\eeq
Here, we define the Euclidean inner product of two discretely sampled time series, $\bm a$ and $\bm b$, as $\bm a\cdot\bm b=\sum_i a(t_i)b(t_i)$, where $t_i$ denotes the $i$th sampling time.  The PSO search optimizes the $10$ source parameters other than the luminosity distance $d_L$  using the local-best PSO implementation in \texttt{PySwarms}~\cite{Kennedy:1995mri,Miranda:2018pyswarms}. The optimization proceeds in successive rounds, each consisting  of $200$ iterations.
At the beginning of each subsequent round, the $40$ particles with
the highest $\mathcal O_{AE}$ from the preceding round are retained.
The search is terminated at the end of a round once a candidate
with $\mathcal O_{AE}>0.9998$ has been found.
Once the PSO search is completed, the luminosity distance $d_L$ is
determined analytically from the candidate parameter values that yield  the highest $\mathcal O_{AE}$ (see Supplemental Material~\cite{SupplementalMaterial} for details of the PSO implementation and numerical settings).

Finally, the posterior distributions of the EMRI source parameters are obtained through a staged MCMC sampling strategy implemented  with \texttt{Eryn}~\cite{Karnesis:2023ras}, adopting  the prior distributions listed in Table~\ref{tab:parameter_summary}. In the first stage, MCMC sampling is performed using a short data segment, with the chains initialized near the parameter values identified by PSO. At each subsequent stage, the data duration is increased, and the  chains are initialized near the  highest-likelihood sample obtained from the preceding stage
(see Supplemental Material~\cite{SupplementalMaterial}
for the likelihood).

\paragraph*{Results---}
\label{sec:results}
The neural network is trained on four NVIDIA Tesla V100 graphics processing units  for  approximately $10$ days. To mitigate overfitting, new EMRI signals are generated at each training epoch, yielding approximately $1.5\times10^7$ training samples in total. The model with the lowest validation loss is selected for the subsequent analyses. Once trained,   the neural network  generates $2000$ approximate posterior samples  for a given EMRI signal in approximately $0.1\,{\rm s}$.

We first  assess  the reduction in the volume of the parameter search
space using an independent dataset comprising $20{,}000$ EMRI
signals, whose source parameters are drawn from the priors
listed in Table~\ref{tab:parameter_summary}.
For each signal in this dataset, the reduction in the search region for each parameter is quantified by the range retention ratio
$r_{ij}=\Delta\theta_{ij,\mathrm{search}}/\Delta\theta_{i,\mathrm{prior}}$,
where $i$ and $j$ label  the source parameter and the signal,
respectively.
Here, $\Delta\theta_{ij,\mathrm{search}}$ denotes the  width of the network-predicted search interval, and
$\Delta\theta_{i,\mathrm{prior}}$ is the corresponding prior width.

The resulting search intervals are substantially narrower than the prior ranges for most parameters.
In particular, the median retention ratios are approximately $0.755\%$ for $e_0$ and
$0.0573\%$ for $d_L$.
In contrast, the medians of $r_{ij}$ are approximately $46.8\%$ for
$\cos\theta_K$ and $30.8\%$ for $\phi_K$, indicating less pronounced contraction
for the MBH spin orientation parameters
(see the upper panel of Fig.~\ref{fig:flow_range_contraction} in the Supplemental
Material~\cite{SupplementalMaterial} for the distributions of
$r_{ij}$).

To quantify the reduction in the overall volume of the $11$-dimensional
parameter search volume, we define the volume contraction factor as
$C_{V,j}=\prod_{i=1}^{11}r_{ij}^{-1}$.
This quantity measures the volume reduction of the axis-aligned search region relative to the
original prior domain, without accounting for correlations among parameters.
Across the $20{,}000$ test signals,  $C_{V,j}$ have a median
of approximately $2.58\times10^{15}$ and a minimum of approximately
$1.17\times10^8$.
In addition, for $98.345\%$ of the test signals, all $11$ true
parameter values lie within their respective search intervals. Together, these results demonstrate that the neural network substantially
reduces the volume of the parameter search space while retaining the true
parameter values for a large majority of the test signals.

Although all three initial phases are fixed to zero during training,
we also evaluate the network's  performance on signals with random initial phases.
For this purpose, we generate a second independent  test dataset of $20{,}000$ EMRI signals, with the
three initial phases independently drawn from
uniform distributions 
over
$[0,2\pi)$.
The resulting  distributions of $r_{ij}$ are comparable to those obtained
for signals with fixed initial phases
(see the lower panel of Fig.~\ref{fig:flow_range_contraction} in the Supplemental
Material~\cite{SupplementalMaterial} for the distributions of
$r_{ij}$).
The median of $C_{V,j}$ is approximately
$2.64\times10^{15}$, compared with
$2.58\times10^{15}$ for the dataset with fixed initial phases.
In addition, for $98.345\%$ of the signals in the second dataset,
all $11$ true parameter values lie within their corresponding
search regions, matching the joint coverage fraction obtained
for the dataset with fixed initial phases.
These results indicate that the network achieves
similar reductions in the volume of the parameter search space
and comparable coverage of the true parameter values for signals
with random initial phases, despite being trained with fixed
initial phases.

We next  assess the complete inference pipeline with a simulated EMRI signal, denoted as  Signal~1. The  source parameters are randomly drawn from their respective prior distributions, with all three initial phases fixed to zero. The signal has a signal-to-noise ratio (SNR) of $95.94$ over a six-month observation period. The first three months of data are then fed into the trained neural network to identify a reduced search region for PSO
(see Fig.~\ref{fig:flow_signal1} and Table~\ref{tab:signal_results}
in the Supplemental Material~\cite{SupplementalMaterial}
for the parameter distributions generated by the network
and the corresponding search regions, respectively).

After $8$ rounds of PSO, the joint  overlap exceeds the
stopping threshold.
The resulting parameter estimates are then used to initialize
the MCMC chains in the first stage
(see Fig.~\ref{fig:pso_final_overlap} in the Supplemental
Material~\cite{SupplementalMaterial} for the overlap evolution
during the PSO search).
At successive MCMC stages, the data duration is progressively
increased, reaching six months in the final stage.
With the neural network pretrained, the complete
inference procedure takes approximately $10\,{\rm h}$.
Figure~\ref{fig:result_high_snr} compares the resulting
posteriors with those obtained from a reference
MCMC analysis with chains initialized near the true parameter values.
Both analyses adopt the same priors and MCMC
settings. The posterior distributions are in close
agreement, exhibiting substantially overlapping
credible regions and similar parameter correlations.

To investigate a less favorable case for neural-network-guided search-space reduction,
we further consider a second signal, denoted as Signal~2, which has the smallest volume contraction factor, $C_{V,j}=1.165\times10^8$, among the $20{,}000$ simulated signals with fixed initial phases. Its SNR over six months is $17.75$ (see Fig.~\ref{fig:flow_signal2} and Table~\ref{tab:signal_results}
in the Supplemental Material~\cite{SupplementalMaterial}
for the parameter distributions generated by the network and the corresponding search regions, respectively). After $11$ rounds of PSO, the joint overlap $\mathcal O_{AE}$ exceeds the stopping threshold, and the resulting parameter values are used to initialize the first-stage MCMC chains.
(see Fig.~\ref{fig:pso_final_overlap} in the Supplemental Material~\cite{SupplementalMaterial} for the overlap evolution during the PSO search). The same staged MCMC procedure used  for Signal~1 is then applied,  yielding posterior distributions in good agreement with those obtained from reference MCMC runs  initialized near the true parameter values (see Fig.~\ref{fig:result_low_snr} in the Supplemental Material~\cite{SupplementalMaterial}). The complete inference pipeline also takes approximately $10\,{\rm h}$, with MCMC sampling dominating  the computational cost. Together, these two examples demonstrate that the proposed framework can efficiently recover posterior distributions consistent with reference MCMC results without requiring knowledge of the true source parameters for initialization.

\paragraph*{Conclusion---}\label{sec:conclusion}

We have developed a deep-learning-assisted Bayesian inference framework for EMRIs that combines neural-network-guided parameter-space localization, particle swarm optimization, and staged MCMC sampling. 
By using the neural network to identify reduced search regions and PSO to locate suitable starting points,
the framework enables subsequent likelihood-based posterior sampling without requiring prior knowledge
of the true source parameters.

Tests on $20{,}000$ simulated EMRI signals show that the neural network substantially reduces
the parameter search volume while maintaining high coverage of the true parameter values.
For signals with fixed initial phases, the median volume contraction factor reaches approximately
$2.58\times10^{15}$, with a joint coverage fraction of $98.345\%$. Comparable performance is obtained for signals with random initial phases, despite the network having been trained exclusively on fixed-phase
signals. These results demonstrate the effectiveness of the network in reducing the search space and its robustness to variations in the initial phases.

For two simulated six-month EMRI signals with SNRs of $95.94$ and $17.75$, the complete inference pipeline yields posterior distributions in close agreement with reference MCMC analyses initialized near the true parameter values. Once the network has been trained, each inference run takes approximately $10\,\mathrm{h}$. These results demonstrate the potential of combining learned parameter-space localization with likelihood-based optimization and sampling to facilitate EMRI Bayesian inference.

The present study is a proof-of-principle demonstration based on noise-free simulations and a restricted source-parameter domain. Realistic EMRI observations will contain instrumental noise and an unresolved Galactic binary foreground and may be affected by glitches, data gaps, and uncertainties or variations in the noise power spectral density. Addressing these effects will require training and validating the network on more realistic simulated data and over wider parameter ranges. Further improvements could include expanding
the network-predicted search regions to increase parameter coverage and adapting the framework
to more sophisticated waveform models. Such developments would be important steps toward applying the proposed approach to EMRI signals in future space-based GW observations.

\begin{acknowledgments}
	\paragraph*{Acknowledgments---}
	This work was supported by the National Natural Science Foundation
	of China (Grants No.~12405056, No.~12275080, No.~12203004, and
	No.~12635002), the Natural Science Foundation of Hunan Province
	(Grants No.~2025JJ40006 and No.~2026JJ60110), and the Hunan Provincial
	Innovation Research Team Project (Grant No.~2024JJ1006).
	\end{acknowledgments}

	\bibliography{references}

\clearpage
\onecolumngrid
\setcounter{figure}{0}
\setcounter{table}{0}
\setcounter{equation}{0}
\renewcommand{\thefigure}{S\arabic{figure}}
\renewcommand{\thetable}{S\arabic{table}}
\renewcommand{\theequation}{S\arabic{equation}}
\renewcommand{\theHfigure}{supp.\arabic{figure}}
\renewcommand{\theHtable}{supp.\arabic{table}}
\renewcommand{\theHequation}{supp.\arabic{equation}}
\section*{Supplemental Material}
\label{supp:material}
\label{app:methods}

\setcounter{secnumdepth}{3}
\setcounter{section}{0}
\renewcommand{\thesection}{\Roman{section}}
\renewcommand{\theHsection}{supp.\arabic{section}}
\makeatletter
\@booleanfalse\raggedcolumn@sw
\@removefromreset{equation}{section}
\def\@sectioncntformat#1{\csname the#1\endcsname.\quad}
\def\@hangfrom@section#1#2#3{\@hangfrom{#1#2}\MakeTextUppercase{#3}}
\makeatother
\raggedbottom
\clubpenalty=10000
\widowpenalty=10000

This Supplemental Material provides additional results and methodological
details for the proposed deep-learning-assisted Bayesian inference framework for extreme-mass-ratio inspirals (EMRIs).
Section~\ref{app:ml} evaluates the ability of the neural network in
reducing the parameter search space, including its robustness to
variations in the initial phases.
Section~\ref{supp:inference} details the complete inference pipeline for  two
test signals, covering neural-network-guided parameter space localization,
particle swarm optimization (PSO), and  Markov chain Monte Carlo (MCMC) posterior sampling.

\section{Neural-Network-Guided  Parameter Space Reduction}
\label{app:ml}

Once trained, the neural network rapidly generates approximate posterior samples of the source parameter  for a given
EMRI signal. For each signal, we generate $2000$ samples and use their $99.73\%$ marginal
credible intervals to construct the reduced search region for PSO. The marginal distributions
of the angular parameters $\cos\theta_S$, $\phi_S$, $\cos\theta_K$, and $\phi_K$ can be
multimodal, resulting in disjoint credible intervals. To accommodate this multimodality, we
compute their highest-density intervals using \texttt{arviz.stats.hdi} with
\texttt{method="multimodal"}.  For each parameter, the retained search range is defined as the union of these intervals, with its total width obtained by summing the widths of the individual intervals, excluding the gaps between them.
We then quantify the reduction in
parameter space using the range retention ratio $r_{ij}$ and the volume contraction factor
$C_{V,j}$ defined in the main text. The latter measures the contraction of the resulting multidimensional search region relative to the original prior domain, without accounting for correlations among parameters.

\begin{figure}[!htbp]
    \centering
    \includegraphics[width=\textwidth,height=0.36\textheight,keepaspectratio]
    {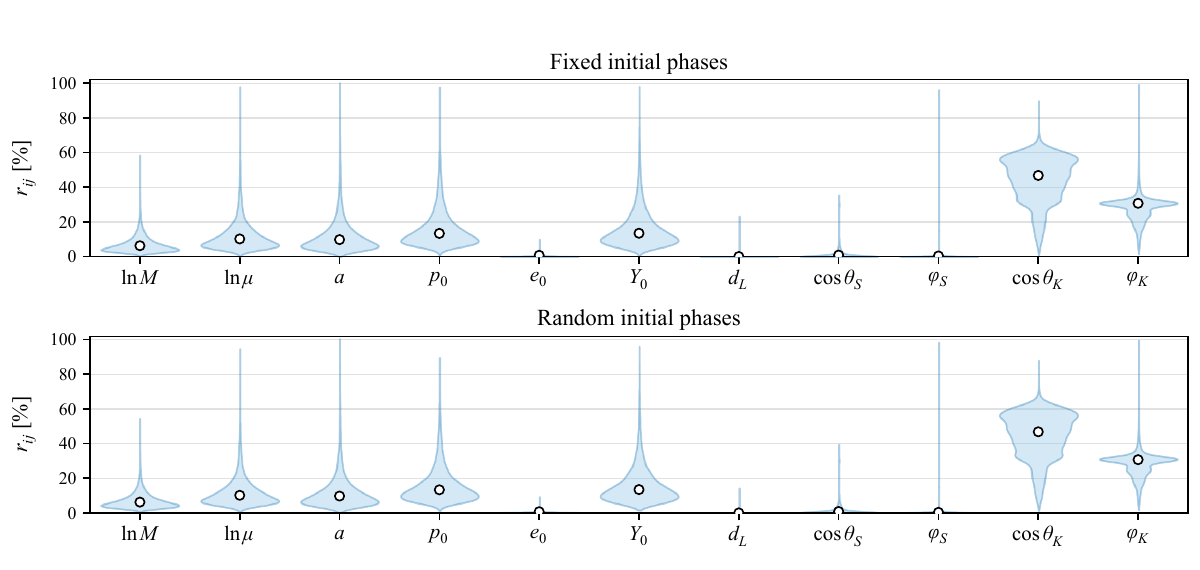}
    \caption{
        Distributions of the range retention ratio $r_{ij}$ for the test
        datasets with fixed and random initial phases.
        The upper panel shows the results for fixed initial phases,
        and the lower panel shows those for random initial phases.
        Each panel includes all $11$ source parameters.
        Blue violins show the full distributions, and open circles mark
        the medians.
        One fixed-phase signal has repeated neural-network samples,
        producing zero widths for some parameter search regions and a zero search
        volume; it is excluded from the $C_{V,j}$ statistics but retained
        in the coverage calculation.
    }
    \label{fig:flow_range_contraction}
\end{figure}

We first evaluate the network on an independent test set of $20{,}000$ EMRI signals drawn
from the same parameter priors as the training data, with the initial phases fixed to
zero. The resulting distributions of $r_{ij}$ are shown in the upper panel of
Fig.~\ref{fig:flow_range_contraction}. The network substantially narrows the search ranges
for most parameters. For example, the median retention ratios are $0.755\%$ for $e_0$ and
$0.0573\%$ for $d_L$. In contrast, the spin orientation parameters $\cos\theta_K$ and
$\phi_K$ exhibit less pronounced contraction, with median ratios of $46.8\%$ and $30.8\%$,
respectively. The resulting volume contraction factor has a median 
$2.58\times10^{15}$ and a minimum of $1.17\times10^8$, with  a central $80\%$ interval of
$[2.57\times10^{13},1.40\times10^{17}]$.

Although all three initial phases are fixed to zero during training, we also evaluate the
network's performance on signals with random initial phases. For this purpose, we generate a second independent test set
of $20{,}000$ EMRI signals, with the three initial phases independently drawn from a
uniform distribution over $[0,2\pi)$. The resulting distributions of $r_{ij}$, shown in
the lower panel of Fig.~\ref{fig:flow_range_contraction}, are comparable to those obtained
for fixed-phase signals. The median retention ratios are $0.751\%$ for $e_0$, $0.0596\%$
for $d_L$, $46.8\%$ for $\cos\theta_K$, and $30.8\%$ for $\phi_K$. The volume contraction
factor $C_{V,j}$ has a median of $2.64\times10^{15}$ and a minimum of $2.66\times10^8$,
with a central $80\%$ interval of $[2.75\times10^{13},1.33\times10^{17}]$. These results
demonstrate that the network achieves comparable reductions in the parameter search space for fixed and random initial phases, indicating its localization performance is relatively insensitive to variations in
the initial phases over the parameter ranges considered.

\par

\section{FULL BAYESIAN INFERENCE PIPELINE}
\label{supp:inference}
\label{app:inference}

To assess the performance of the full inference pipeline, we analyze the two simulated six-month EMRI
signals introduced in the main text (Signal~1 and Signal~2) . Signal~1  has source parameters 
randomly drawn from the training priors and has  a signal-to-noise ratio (SNR) of $95.94$, whereas Signal~2 has the
smallest volume contraction factor $C_{V,j}$ among the $20{,}000$ fixed-phase test
signals, with an SNR of $17.75$. The true source parameters for both signals are listed in
Table~\ref{tab:signal_results}. For each signal, the neural network first identifies 
reduced parameter search regions, PSO then locates high-overlap candidates within this region, and
staged MCMC sampling is subsequently used to obtain the final posterior distributions.

\newlength{\suppbodyparindent}
\setlength{\suppbodyparindent}{\parindent}
\subsection*{Neural-Network-Guided Parameter Space Localization}
\label{app:flow}

For each signal, the first three months of data are fed into the trained neural network,
which generates $50{,}000$ source parameter samples in approximately $3\,\mathrm{s}$.
Following the procedure described in Sec.~\ref{app:ml}, the $99.73\%$ marginal credible
intervals of these samples are used to construct the reduced search region for PSO.

\begin{figure}[!htbp]
    \centering
    \includegraphics[width=0.96\textwidth]{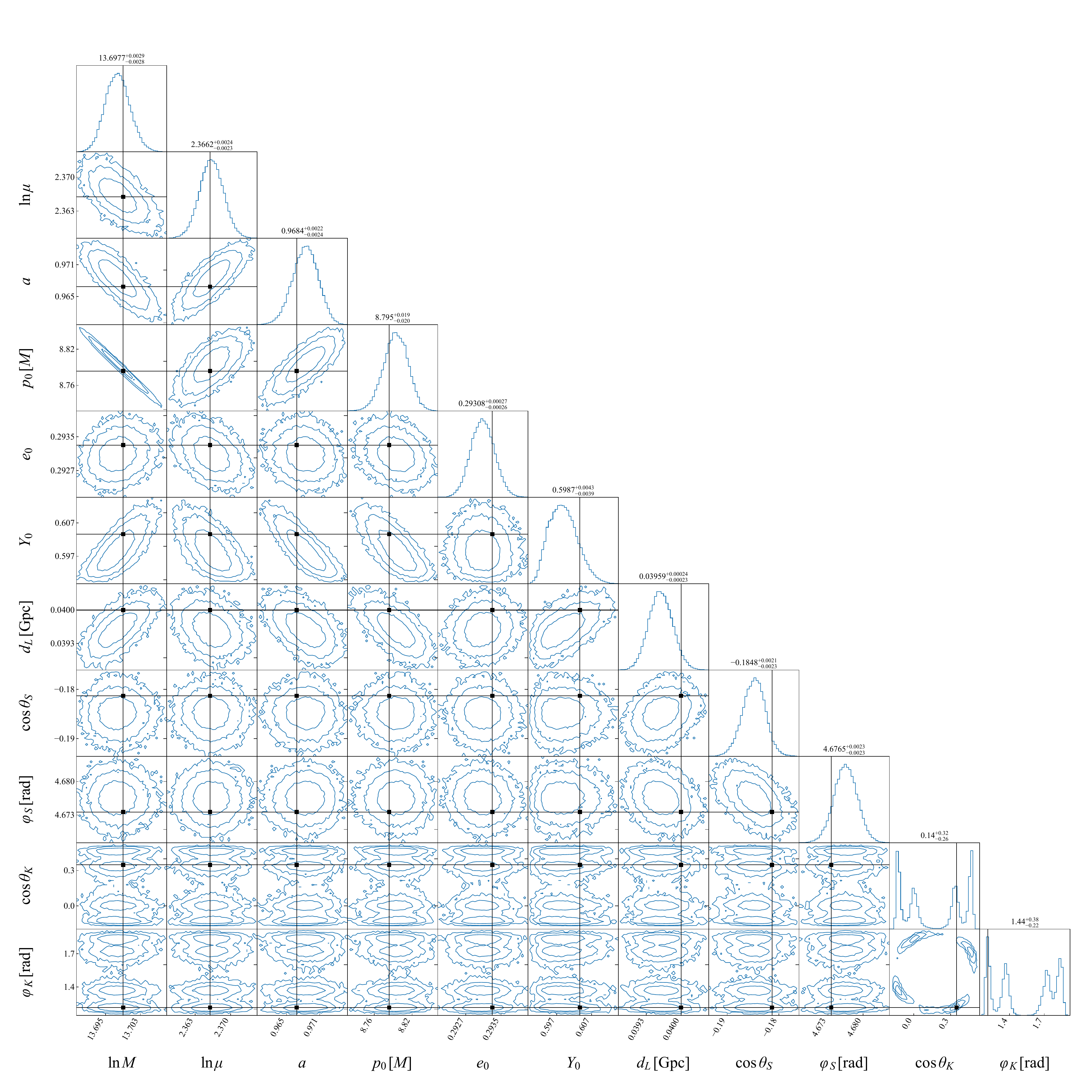}
    \captionof{figure}{
        Corner plot of the network-generated parameter distributions for
        Signal~1, using $50{,}000$ samples.
        The medians and central $68.27\%$ credible intervals are given
        above the diagonal.
        The contours show the $68.27\%$, $95.45\%$, and $99.73\%$
        credible regions.
        Black lines and squares indicate the true values.
    }
    \label{fig:flow_signal1}
\end{figure}

Figure~\ref{fig:flow_signal1} shows the network-generated parameter distributions for Signal~1, with the corresponding search intervals listed in
Table~\ref{tab:signal_results}. The distributions are relatively narrow for most
parameters, with multimodality appearing only in the spin orientation parameters
$\cos\theta_K$ and $\phi_K$. The resulting volume contraction factor is $C_{V,j}=
2.75\times10^{17}$, and all $11$ true parameter values lie within their respective
search intervals. Thus, for Signal~1, the network substantially reduces the parameter search space while
retaining the true source  parameters.

\par\medskip
\par

\begin{figure}[!htbp]
    \centering
    \includegraphics[width=0.96\textwidth]{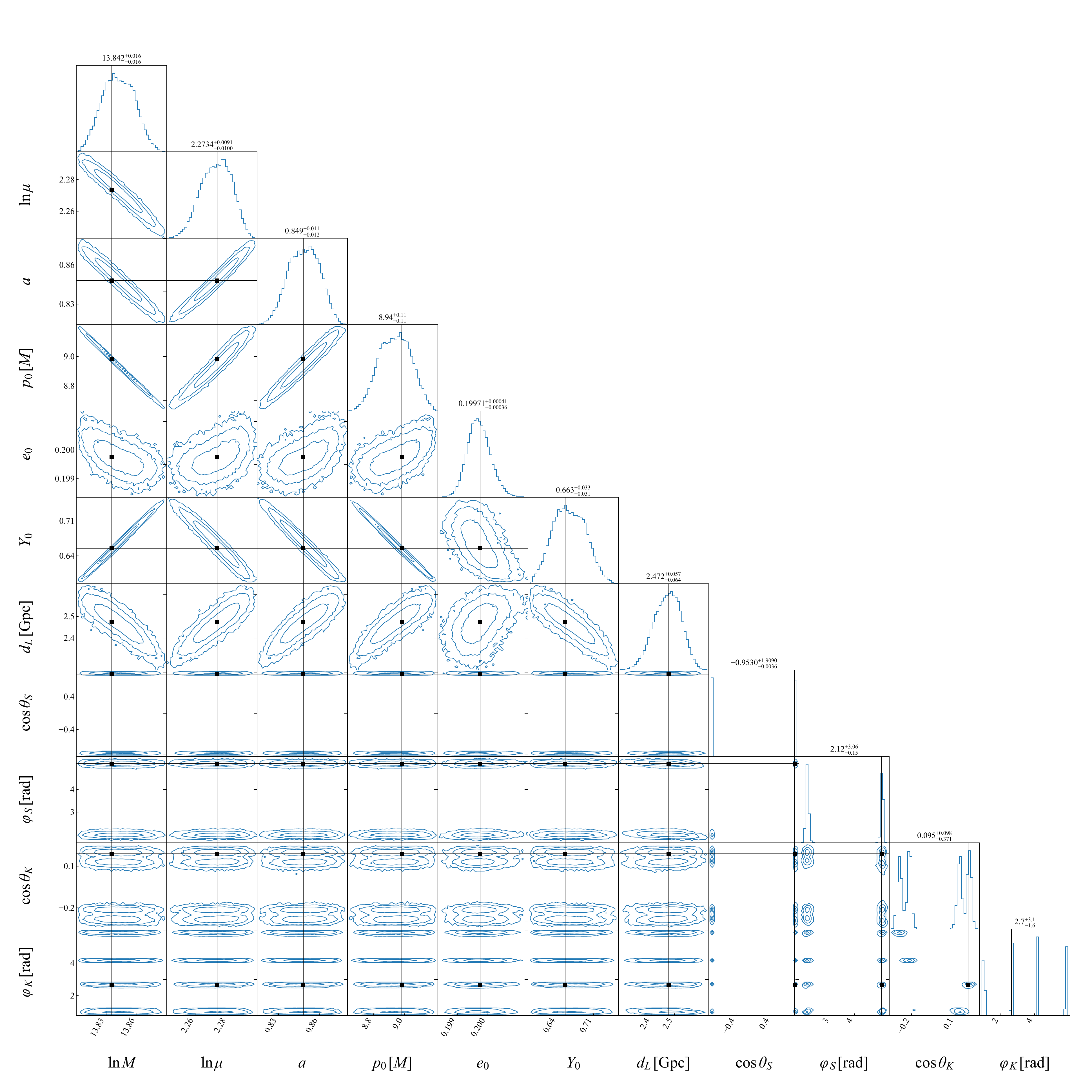}
    \captionof{figure}{
        Corner plot of the network-generated parameter distributions for
        Signal~2, using $50{,}000$ samples.
        The medians and central $68.27\%$ credible intervals are given
        above the diagonal.
        The contours show the $68.27\%$, $95.45\%$, and $99.73\%$
        credible regions.
        Black lines and squares indicate the true parameter values.
        Compared with Signal~1, several marginal distributions are broader,
        and additional modes appear in the
        sky location parameters.}
    \label{fig:flow_signal2}
\end{figure}

The corresponding results for Signal~2 are shown in Fig.~\ref{fig:flow_signal2}, with the
search intervals given in Table~\ref{tab:signal_results}. Compared with Signal~1, the  network-generated
parameter distributions are generally broader, with
additional multimodal structures appearing in the sky location parameters $\cos\theta_S$
and $\phi_S$, alongside the spin orientation parameters $\cos\theta_K$ and $\phi_K$. The volume contraction factor is $C_{V,j}=
1.165\times10^8$, considerably smaller than that for Signal~1. Nevertheless, all $11$ true parameter values remain within their
respective search intervals.  The network therefore retains the true parameters while providing a substantial,
albeit less pronounced, reduction in the search space for Signal~2.

\par\medskip
\par\medskip

\begin{table}[!htbp]
    \centering
    \captionof{table}{
True parameter values, search regions, and PSO results for the
        two EMRI test signals.
    }
    \label{tab:signal_results}
    \small
    \setlength{\tabcolsep}{5pt}
    \renewcommand{\arraystretch}{1.40}
    \begin{tabular*}{\textwidth}{@{\extracolsep{\fill}}lccc}
        \hline\hline
        \multicolumn{4}{c}{Signal~1} \\
        \hline
        Parameter & True value & Search region & PSO result \\
        \hline
        $\ln(M/M_\odot)$ & $13.6990$ & $[13.6900,13.7069]$ & $13.6979$ \\
        $\ln(\mu/M_\odot)$ & $2.3660$ & $[2.3591,2.3735]$ & $2.3660$ \\
        $a$ & $0.9668$ & $[0.9612,0.9748]$ & $0.9679$ \\
        $p_0\,[M]$ & $8.7837$ & $[8.7319,8.8491]$ & $8.7913$ \\
        $e_0$ & $0.2933$ & $[0.2923,0.2939]$ & $0.2933$ \\
        $Y_0$ & $0.6036$ & $[0.5903,0.6120]$ & $0.6016$ \\
        $\cos\theta_S$ & $-0.1813$ & $[-0.1921,-0.1775]$ & $-0.1838$ \\
        $\phi_S\,[\mathrm{rad}]$ & $4.6736$ & $[4.6679,4.6855]$ & $4.6728$ \\
        $\cos\theta_K$ & $0.3486$ & $[-0.2041,0.1760]\cup[0.1909,0.5449]$ & $0.3417$ \\
        $\phi_K\,[\mathrm{rad}]$ & $1.2133$ & $[1.1259,1.4864]\cup[1.5785,1.9297]$ & $1.2119$ \\
        $d_L\,[\mathrm{Gpc}]$ & $4.0000\times10^{-2}$ & $[3.8903,4.0366]\times10^{-2}$ & $3.9846\times10^{-2}$ \\
        \hline
        \multicolumn{4}{c}{Signal~2} \\
        \hline
        Parameter & True value & Search region & PSO result \\
        \hline
        $\ln(M/M_\odot)$ & $13.8354$ & $[13.8057,13.8812]$ & $13.8276$ \\
        $\ln(\mu/M_\odot)$ & $2.2734$ & $[2.2464,2.2943]$ & $2.2786$ \\
        $a$ & $0.8481$ & $[0.8190,0.8744]$ & $0.8542$ \\
        $p_0\,[M]$ & $8.9836$ & $[8.6809,9.1847]$ & $9.0360$ \\
        $e_0$ & $0.1998$ & $[0.1986,0.2011]$ & $0.2001$ \\
        $Y_0$ & $0.6555$ & $[0.5948,0.7424]$ & $0.6404$ \\
        $\cos\theta_S$ & $0.9560$ & $[-0.99999,-0.8169]\cup[0.8164,0.99999]$ & $0.9582$ \\
        $\phi_S\,[\mathrm{rad}]$ & $5.1421$ & $[1.7012,2.3002]\cup[4.8712,5.4418]$ & $5.1618$ \\
        $\cos\theta_K$ & $0.1889$ & $[-0.4109,-0.0874]\cup[-0.0022,0.3247]$ & $0.1851$ \\[3pt]
        $\phi_K\,[\mathrm{rad}]$ & $2.6571$ & $\begin{gathered}
            {}[0.8207,1.3047]\cup[2.4183,2.8969]\\[-3pt]
            {}\cup[3.9408,4.4232]\cup[5.5737,6.0516]
          \end{gathered}$ & $2.6510$ \\[3pt]
        $d_L\,[\mathrm{Gpc}]$ & $2.4748$ & $[2.2995,2.6183]$ & $2.5003$ \\
        \hline\hline
    \end{tabular*}
\end{table}

\subsection*{Particle Swarm Optimization}
\label{app:pso}
Following neural-network-guided localization, PSO is used to identify high-overlap
candidates, with a subset of the particles initialized using network-generated samples.
The optimization is performed using $14$ days of data and seeks  to maximize the joint $A/E$ overlap
$\mathcal O_{AE}$ defined in the main text. Since the normalized overlap is independent of the
overall waveform amplitude, the optimization involves only the 10 source parameters other
than the luminosity distance $d_L$. The PSO search proceeds in successive rounds, each
consisting of $200$ particles and $200$ iterations. In the first round, $100$ particles
are initialized using  network-generated samples, while the remaining 100 are
randomly drawn from the  reduced search region. In each subsequent round, $100$ particles are initialized
from network-generated  samples, 60 are drawn from the same search region, and the $40$
particles with the highest overlaps from the preceding round are retained. The PSO control
parameters~\cite{Miranda:2018pyswarms} follow predetermined cosine-annealing schedules,
and the optimization terminates at the end of a round once a candidate satisfying $\mathcal O_{AE}>0.9998$ has been identified.

\begin{samepage}

After the PSO search, the luminosity distance $d_L$ is determined analytically for  the candidate yielding the highest overlap. Since the waveform amplitude scales inversely with luminosity distance, the best-fit distance
for fixed values of the other source parameters is given by

\end{samepage}
\beq{pso_distance}
d_L
=
d_{L,\mathrm{ref}}
\frac{
\sum_{I=A,E}
\bm h_{I,\mathrm{ref}}\!\cdot\!\bm h_{I,\mathrm{ref}}
}{
\sum_{I=A,E}
\bm d_I\!\cdot\!\bm h_{I,\mathrm{ref}}
},
\eeq
where $\bm h_{I,\mathrm{ref}}$ is the waveform generated using the optimized values of the
other source parameters and a reference luminosity distance
$d_{L,\mathrm{ref}}=1\,\mathrm{Gpc}$. The resulting 11-dimensional parameter vector is
then used to initialize the first-stage MCMC chains.

\phantomsection
\label{app:examples}

\begin{figure}[!htbp]
    \centering
    \includegraphics[width=0.50\textwidth]{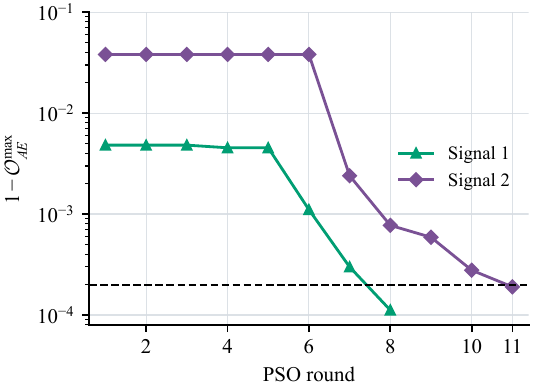}
    \captionof{figure}{
        Plot of the best joint $A/E$ overlap during PSO for Signal~1
        (green triangles) and Signal~2 (purple diamonds).
        For clarity, $1-\mathcal O_{AE}^{\max}$ is shown on a logarithmic
        vertical axis.
        The dashed line at $2\times10^{-4}$ corresponds to the stopping threshold
        $\mathcal O_{AE}=0.9998$.
        The stopping criterion is satisfied after $8$ and $11$ rounds,
        respectively, and each curve ends at that round.
    }
    \label{fig:pso_final_overlap}
\end{figure}

Figure~\ref{fig:pso_final_overlap} shows the evolution of the best joint $A/E$ overlap during PSO for both signals. Here, $\mathcal O_{AE}^{\max}$ denotes the highest overlap achieved
up to each round. For Signal~1, the overlap improves only slightly during the first five
rounds, followed by a more rapid increase until the stopping criterion is satisfied after
eight rounds. Signal~2 exhibits a similar trend, with limited  improvement during the first
six rounds before reaching the stopping threshold after $11$ rounds. The larger number of
rounds required for Signal~2 is consistent with its broader search region. The final PSO
parameter estimates for both signals are listed in Table~\ref{tab:signal_results}.

\par

\subsection*{MCMC Sampling}
\label{app:mcmc}

The posterior distributions of the EMRI source parameters are obtained
through staged MCMC sampling.
For the $A$ and $E$ channels, the likelihood is constructed using the
noise-weighted inner product.
For two signals $\bm a_I$ and $\bm b_I$ in channel $I\in\{A,E\}$, this inner
product is defined as

\beq{mcmc_inner_product}
(\bm a_I|\bm b_I)
=
4\,\mathrm{Re}
\int_0^\infty
\frac{\tilde a_I(f)\tilde b_I^*(f)}
     {S_{n,I}(f)}
\,\mathrm{d}f,
\eeq
where a tilde denotes a Fourier transform, an asterisk denotes complex conjugation, and
$S_{n,I}(f)$ is the one-sided  Laser Interferometer Space Antenna (LISA) noise power spectral density in channel $I$. Assuming
stationary Gaussian noise, the likelihood is given by~\cite{Speri:2024qak}

\beq{mcmc_likelihood}
\mathcal{L}(\bm d|\bm\theta)
\propto
\exp\!\left[
-\frac{1}{2}
\sum_{I=A,E}
\left(
\bm d_I-\bm h_I(\bm\theta)
\mid
\bm d_I-\bm h_I(\bm\theta)
\right)_I
\right].
\eeq
Although the simulated signals contain no
added noise, the likelihood uses the assumed
LISA noise power spectral density to weight
the waveform residuals.

We perform MCMC sampling using \texttt{Eryn}~\cite{Karnesis:2023ras}, with $26$ walkers at
each stage. The first stage analyzes a short data segment, with the walkers initialized
near the parameter values identified by PSO through small random perturbations. At each
subsequent stage, the data duration is increased, and the walkers are initialized near the
highest-likelihood sample from the preceding stage.

\par

\begin{figure}[!htbp]
    \centering
    \includegraphics[width=0.96\textwidth]{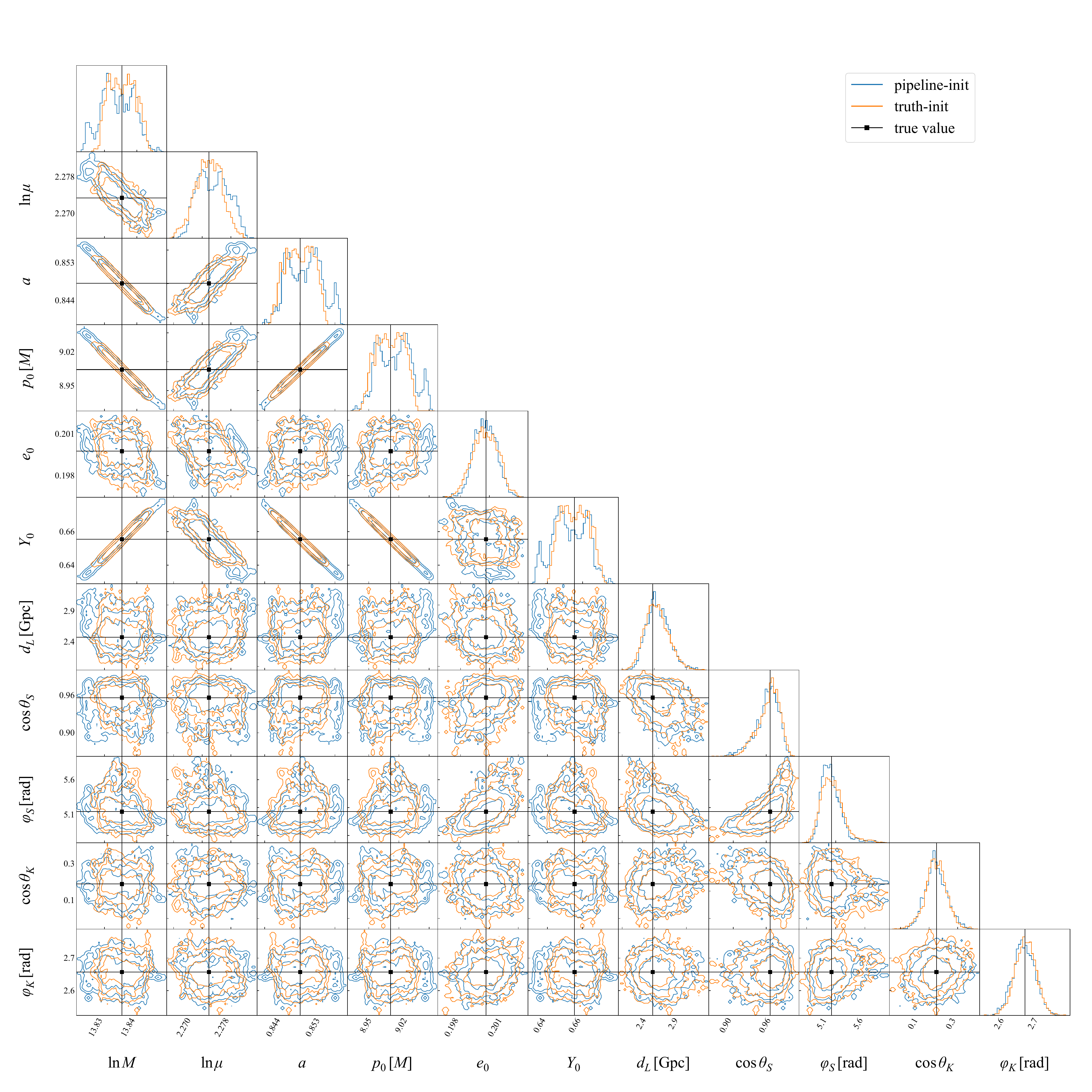}
    \captionof{figure}{
        Posterior comparison plot for Signal~2 using $6$ months of
        data, with an SNR of $17.75$.
        Posteriors obtained through MCMC analyses with
        progressively longer data segments are shown in blue, with the
        chains in the first stage initialized near the parameter values identified by the PSO search
        assisted by the neural network.
        The posterior obtained with MCMC chains initialized near
        the true parameter values is shown in orange.
        The contours show the $68.27\%$, $95.45\%$, and $99.73\%$
        credible regions.
        Identical priors and MCMC settings are used in both analyses.
        True values are marked by black lines and squares.
    }
    \label{fig:result_low_snr}
\end{figure}

To evaluate the performance of the complete inference pipeline, we compare the resulting six-month posteriors with those obtained from reference MCMC analyses initialized near the true source
parameters. The comparison for Signal~1 is presented in the main text, while
the corresponding results for Signal~2 are shown in Fig.~\ref{fig:result_low_snr}.  For Signal~2, 
the marginal posteriors of $\ln(M/M_\odot)$, $\ln(\mu/M_\odot)$, $a$, $p_0$ and $Y_0$
exhibit some differences in their widths or weak secondary structures between the two
analyses. The remaining parameters show closer agreement, with substantially overlapping
credible regions and similar parameter correlations. Overall, the proposed pipeline yields 
posterior distributions broadly consistent with the reference MCMC results, despite the
less pronounced search-space reduction achieved by the neural network for Signal~2.

\par

\end{document}